\documentstyle[12pt]{article}

\def\be{\begin{equation}}
\def\ee{\end{equation}}
\def\bdi{\begin{displaymath}}
\def\edi{\end{displaymath}}
\def\br{\begin{eqnarray}}
\def\er{\end{eqnarray}}
\def\no{\nonumber}
\def\o{\over}

\def\u2{\mid u\mid^2}

\def\ra{\rightarrow}
\def\no{\nonumber}

\def\RR{{\rm I\kern-.1567em R}}                              
 \def\CC{{\rm C\kern-4.7pt                                    
 \vrule height 7.7pt width 0.4pt depth -0.5pt \phantom {.}}} 
 \def\ZZ{{\sf Z\kern-4.5pt Z}}                                

\begin{document}

\begin{titlepage}
\vspace*{-2 cm}
\noindent

\vskip 3cm
\begin{center}
{\Large\bf Soliton stability in some knot soliton models  }
\vglue 1  true cm

C. Adam$^{a*}$,   J. S\'anchez-Guill\'en$^{a**}$,  
and A. Wereszczy\'nski$^{b\dagger}$
\vspace{1 cm}

\small{ $^{a)}$Departamento de Fisica de Particulas, Universidad
      de Santiago}
      \\
      \small{ and Instituto Galego de Fisica de Altas Enerxias (IGFAE)}
     \\ \small{E-15782 Santiago de Compostela, Spain}
      \\ \small{ $^{b)}$Institute of Physics,  Jagiellonian
     University,}
     \\ \small{ Reymonta 4, 30-059 Krak\'{o}w, Poland}

\medskip
\end{center}

\normalsize
\vskip 0.2cm

\begin{abstract}
We study the issue of stability of static soliton-like solutions in some
non-linear field theories which allow for knotted field configurations.
Concretely, we investigate the AFZ model,
 based on a Lagrangian 
quartic in first derivatives with infinitely many
conserved currents, for which infinitely many soliton solutions are known
analytically. For this model we find that sectors with different
(integer) topological charge (Hopf index) are {\em not} separated by an
infinite energy barrier. Further, if variations which change the topological
charge are allowed, then the static solutions are not
even critical points of the energy functional. We also explain why soliton
solutions can exist at all, in spite of these facts.
In addition, we briefly discuss  the Nicole model, which is
based on a sigma-model type  Lagrangian.
For the Nicole model we find
that different topological sectors are separated by an infinite energy barrier.
\end{abstract}

\vfill

{\footnotesize
$^*$adam@fpaxp1.usc.es

$^{**}$joaquin@fpaxp1.usc.es

$^{\dagger}$wereszczynski@th.if.uj.edu.pl }

\end{titlepage}

\section{Introduction }

Nonlinear field theories, which allow for static soliton
solutions of their field equations, have been studied intensively for
many years due to their importance in a variety of fields in theoretical
and mathematical physics.   Indeed,  applications range from elementary
particle theory, where they may serve as effective field theories
providing particle-like solutions, to condensed matter and solid state
physics. One important criterion in the classification of these
nonlinear field theories is given by the dimension of the base space
(i.e., space-time) on which these fields exist. In the case of one
space and one time dimension, static solutions have to solve a
nonlinear ODE, which is usually much easier to solve than the
nonlinear PDEs which result for solitons in higher dimensions.
In addition, in $1+1$ dimensions many rigorous results on nonlinear
field theories and their soliton solutions are known, and a vast
mathematical apparatus for the analysis of these models has been
developed. Among those are inverse scattering, integrability, and
the zero curvature representation, where the latter generalizes 
to field theory the
Lax pair representation of integrable systems in $0+1$ dimensions
(that is, systems with finitely many degrees of freedom).  

In higher dimensions (i.e., in $d+1$ dimensional space-time for $d>1$),
on the other hand, much less is known about nonlinear field theories
and their solutions.  A generalization of the zero curvature representation
to higher dimensions was proposed in \cite{AFSG}, and it was shown there
that this proposal leads to nonlinear field theories with infinitely
many conservation laws, realizing thereby the concept of integrability
in higher dimensions in a concrete and well-defined manner.
Besides this, some specific nonlinear field theories in higher
dimensions have been studied
with the intention of more phenomenological applications. 
As far as static finite-energy solutions (solitons) are relavant in
these applications, the selected nonlinear theories should, of course, really 
support such static solutions. One necessary condition for the
existence of static finite-energy solutions is provided by the Derrick
scaling argument (see, e.g., \cite{Mak,mansut}). 
The Derrick scaling argument simply says that static
solutions with finite, nonzero energy cannot exist if it is possible for
arbitrary field configurations with finite, nonzero energy to rescale the
energy functional to arbitrarily small values by a base space 
scale transformation ${\bf r} \to \mu {\bf r}$, where ${\bf r}\in \RR^d$
are the base space coordinates
and $\mu$, with $0<\mu<\infty$, is the scale parameter. 
Let us be more precise for the class of theories we are interested in.
We assume that we have only scalar fields $\Phi_a$, $a=1 \ldots N$,
which transform trivially under the above scale transformation, 
$\Phi_a \to \Phi_a$, and that the energy density for static configurations
is a sum of terms each of which is homogeneous (of degree $h$)
in space derivatives $\nabla_k \Phi_a$. 
Then each such term ${\cal E}_h(\Phi_a , \nabla_k \Phi_a)$ contributes
a term $E_h[\Phi_a]= \int d^d{\bf r}{\cal E}_h(\Phi_a , \nabla_k \Phi_a)$ 
to the energy functional, where $E_h[\Phi_a]$
transforms as $E_h[\Phi_a]\to \mu^{d-h}E_h[\Phi_a]$ under the scale
transformation. There are now several possibilities to obey the
Derrick criterion for the possible existence of solutions. If the
energy density only consists of one term, then necessarily $d=h$, that is,
the degree of homogeneity in first derivatives, $h$, must be equal to
the dimension of space, $d$. If the energy density consists of several
terms, then at least one term must obey $d-h>0$, and at least one further term
must have the
opposite behaviour, $d-h<0$ (unless all terms obey the scale invariance
condition $d=h$). All our concrete investigations will be for $3+1$ 
dimensional space-time, therefore we assume $d=3$ from now on.
 
One well-known example of a field theory in $d=3$ dimensions is the 
Skyrme model \cite{Skyrme}, 
which has applications in nuclear physics. It has the group SU(2) as 
field configuration space (target space),\footnote{In particle physics,
SU(3) has to be used, see e.g. \cite{Witt}.}
and the fields are supposed
to describe  pions, whereas the soliton solutions 
are related to the nucleons. The energy functional of the Skyrme model
consists of a sigma-model type term quadratic in first derivatives,
with $d-h =3-2=1$, and of a further term quartic in derivatives
(the ``Skyrme term'') with $d-h=3-4=-1$, so it obeys the Derrick criterion
for the possibility of static solutions. Occasionally a third 
``potential'' term is
added to the Lagrangian density (and, consequently, to the energy density), 
which only depends on the fields, but not on derivatives. This term
is usually assumed to make the pions massive. Soliton solutions have been
found numerically for the Skyrme model both with and without the
pion mass term \cite{HMS,BS1,HoMa}. 

Another well-known model with applications both to field theory and condensed
matter physics is the Faddeev--Niemi model (\cite{Fad}, \cite{FN1}). 
This model may be derived 
from the Skyrme model by 
simply restricting its target space from SU(2) $\sim S^3$ to the 
two-sphere $S^2$ (e.g., the equator of the Skyrme $S^3$, or SU(2)/U(1)). 
The target 
space of the Faddeev--Niemi model is two-dimensional, therefore the
solitons are now line-like instead of point-like (where the soliton
position may be defined, e.g., by the loci where the fields take values
which are antipodal to their vacuum value). More precisely, the solitons
of the Fadeev--Niemi model are links or knots. The corresponding
topological index classifying a finite energy field configuration is a 
linking number (the Hopf index), whereas it is a winding number for the
Skyrme model. The existence of soliton solutions in the Faddeev--Niemi
model has been proven in \cite{LiYa}, and 
confirmed by numerical calculations, e.g., in
(\cite{GH} -- \cite{HiSa}).

The Faddeev--Niemi model has the Lagrangian density
\be \label{FN-L}
{\cal L} = \frac{m^2}{2} {\cal L}_2 - \lambda {\cal L}_4 + {\cal L}_0
\ee
where $m$ is a constant with dimension of mass,
$\lambda $ is a dimensionless coupling constant, 
${\cal L}_2$ is 
\be \label{cp1}
{\cal L}_2 =  \frac{\partial_\mu u \, \partial^\mu \bar u}{(1+ u\bar u)^2} ,
\ee
and ${\cal L}_4$ is
\be
 {\cal L}_4 =  \frac{(\partial^\mu u \, \partial_\mu \bar u)^2 - (\partial^\mu
u \, \partial_\mu u)(\partial^\nu \bar u \, 
\partial_\nu \bar u)}{(1+u\bar u)^4} .
\ee
Further, $u$ is a complex field
which parametrizes the stereographic projection of the target $S^2$, see 
Section 2 below for more details. Again we have allowed for a pure potential
term ${\cal L}_0 (u ,\bar u)$. This term is not required by the Derrick
criterion, but it may be necessary for some applications of the model
(e.g., by providing a symmetry breaking of the target space symmetry
group from SU(2) to U(1)).

It is also possible to select Lagrangian densities which consist of one
term only and are constructed from ${\cal L}_4$ or ${\cal L}_2$ exclusively,
and obey the Derrick criterion $h=3$. They are, however, necessarily
non-polynomial.  
For ${\cal L}_4$ the appropriate choice is
\be
{\cal L}_{\rm AFZ} = -({\cal L}_4)^\frac{3}{4} .
\ee
This model has been introduced and studied by Aratyn, Ferreira and
Zimerman (AFZ) in \cite{AFZ1}, \cite{AFZ2}. 
AFZ found
infinitely many analytic soliton solutions for this model
by using an ansatz with toroidal coordinates.
The analysis of the AFZ model was carried further in 
(\cite{BF}), where, among other results, all the space-time and (geometric) 
target space symmetries of the AFZ model were determined, and, further,
the use of the ansatz with toroidal coordinates was related to the
conformal symmetry of the model (more precisely, of the static equations
of motion).  It turns out that
the AFZ model has infinitely many target space symmetries and, thus,
infinitely many conservation laws. Moreover, it
also shows classical integrability in a different sense,
because the static field equations 
resulting from the ansatz with toroidal
coordinates may be solved by simple integration.

The other model, based solely on ${\cal L}_2$, is the Nicole model
 \be \label{Ni-La}
{\cal L}_{\rm Ni}= ({\cal L}_2)^\frac{3}{2}.
\ee
This model has first been proposed by Nicole (\cite{Ni}), and it was shown
in the same paper that the simplest Hopf map with Hopf index 1 is a
soliton solution for this model.
(This model is, in fact, the 
restriction to $S^2$ target space of a non-polynomial $SU(2)$ model
which was studied first in \cite{Deser} as a possible candidate for
a pion model.)
The Nicole model shares the conformal symmetry with the AFZ model and,
therefore,
the ansatz with toroidal coordinates may be used again to simplify the
static field equations (to reduce them to an ordinary differential equation). 
However, the Nicole model only has the obvious symmetries - the conformal
base space symmetries (in the static case) and the modular target space
symmetries, see \cite{ASG2}. Consequently, the field equations are no longer 
integrable, and the solutions are no longer available in closed,
analytic form (except for the simplest case with Hopf index 
one\footnote{Exact solitons with higher Hopf index have
been found in modified Nicole models \cite{nicole_modif}}). 
For a detailed investigation of soliton solutions within the ansatz
in toroidal coordinates we refer to \cite{ASGVW}.

It is the main purpose of this paper to study stability issues of these
two latter models with explicit solutions. 
Firstly, let us emphasize that the Derrick criterion
is a necessary condition for the existence of solutions, but by no
means a sufficient one. So both the existence and the stability
of soliton solutions have to be investigated independently. 
We find the more surprising results for the AFZ model.\footnote{Warnings about
its stability, based on the base space scale invariance ${\bf r}
\to \mu {\bf r}$ of the model, have been given in \cite{mansut}.}
In this model
it turns out that there exists a symmetry transformation (which maps
solutions into solutions) which connects an arbitrary solution to the
trivial vacuum solution $u=0$. Consequently, soliton solutions with a
nonzero topological index are not separated from the vacuum by an energy
barrier. Further, soliton solutions are not even critical points of the
energy functional, because there exists a nonzero first variation of
the energy functional into the direction of the symmetry transformation.
This immediately poses an apparent paradox: (static) soliton solutions 
solve the variational equations of the energy functional, therefore they
should be critical points of the energy functional by definition.
If, on the other hand, the above-mentioned symmetry implies that there
exists a non-flat direction of the energy functional for arbitrary fields,
this seems to imply that finite energy solutions cannot exist, much like
the original Derrick scaling argument. Static soliton solutions of the
AFZ model do exist, however, and have been constructed in
(\cite{AFZ1}, \cite{AFZ2}). In this paper, therefore, we not only
show the existence of the symmetry transformation connecting
solutions to the vacuum, we also explain how the above-mentioned
apparent paradox is resolved, by considering inherent source terms.

In Section 2 we study the reduced energy functional which follows 
from an ansatz using toroidal coordinates.
In Subsection 2.1 we perform this study for a simple toy model. The toy model
has the same stability problems as the AFZ model, but much simpler
field equations, therefore we mainly introduce it for illustrative
purposes. This toy model may, however, be of some independent interest.
It consists of a non-quadratic kinetic energy term, and such models have
been studied in the context of ``fake inflation'' 
\cite{fake} (where the change
in light propagation caused by  the non-conventional kinetic term 
mimics a nontrivial geometry, as far as the propagation of light is 
concerned), and in the context of the so-called modified Newtonian
dynamics (MOND), see, e.g., \cite{MOND}. 
In addition, the toy model is equivalent
to the electrostatic sector of a nonlinear theory of electrodynamics.
This issue we discuss briefly in Subsection 3.1.
In Subsection 2.2, at first we discuss the boundary conditions implied by the
Hopf index for the ansatz within toroidal coordinates. Then
we study the reduced energy functional of the AFZ
model, using the analogy with the toy model. We establish the instability
and show that the existence of solutions which are not critical points
can be attributed to boundary term contributions to the reduced energy
functional. In Section 2.3 we briefly study the reduced energy functional
of the Nicole model. We find that in this case different topological
sectors are indeed separated by an infinite energy barrier, as one generally
expects for topological soliton models.

In Section 3 we study the full energy functional. In Subsection 3.1 we again
investigate the toy model. We show that the apparent paradox is resolved
by the presence of singular, delta-function like flux terms in a
conservation equation. Further, we briefly introduce the above-mentioned
nonlinear electrodynamics. In Subsection 3.2 we investigate the AFZ
model. It turns out that the apparent paradox related with the existence of
solutions is resolved in this case by the existence of an inhomogeneous,
singular flux term in just one of the infinitely many conservation
equations of the model.

Section 4 is devoted to the discussion of our results, emphasizing both 
the stability problems and the resolution of the apparent paradox. Further,
we discuss possible generalizations of our results, and their application to
and implications for further models.
In Appendix A we calculate the Hopf index for a field within the ansatz
in toroidal coordinates.   

\section{The reduced energy functional}
\subsection{A toy model}
Firstly, let us introduce a simple toy model which already shows the essential
features we want to study in the sequel. We choose the energy functional
\be
E[\Phi ] = \int d^3 {\bf r} (\nabla \Phi \cdot \nabla \Phi )^\frac{3}{2}
\ee
where $\Phi$ is a real scalar field. Here the non-integer power of
$\frac{3}{2}$ is chosen precisely as to make the energy functional
invariant under a base space scale transformation ${\bf r} \to
\mu {\bf r}$, avoiding thereby the usual Derrick scaling instability.
There is, however, another scale transformation which does {\em not} leave 
invariant the
energy functional. In fact, as both the energy functional and the
resulting Euler--Lagrange equations are homogeneous in the scalar field   
$\Phi$, the target space scale transformation $\Phi \to \lambda \Phi$ is
a symmetry of the Euler--Lagrange equations which does not leave 
invariant the energy.
As long as there exists no normalization condition for $\Phi$
which fixes the value of the scale factor $\lambda$ (and enters the
energy functional, e.g., via a Lagrange multiplier), the same Derrick type
scaling argument applies and seems to prevent the existence of finite
energy solutions to the Euler--Lagrange equations. Nevertheless, finite
energy solutions exist, as we want to show now. We introduce toroidal
coordinates $(\eta ,\xi ,\varphi)$, $\eta \in [0,\infty ]$, $\xi ,\varphi
\in [0,2\pi ]$, via
\br
x &=&  q^{-1} \sinh \eta \cos \varphi \;\;, \;\;
y =  q^{-1} \sinh \eta \sin \varphi   \nonumber \\
z &=&  q^{-1} \sin \xi \quad ;  \qquad  q = \cosh \eta - \cos \xi 
\label{tordefs}
\er
in order to reduce the Euler--Lagrange equations to an ordinary 
differential equation (ODE). Further,
we need the volume form
\be
dV \equiv d^3 r = q^{-3} \sinh \eta \, d \eta \, d\xi \, d\varphi
\ee
and the gradient
\be \label{grad-3}
\nabla = (\nabla \eta)\partial_\eta 
+(\nabla \xi )\partial_\xi +(\nabla \varphi)\partial_\varphi
= q(\hat e_\eta \partial_\eta + \hat e_\xi \partial_\xi +
\frac{1}{\sinh \eta} \hat
e_\varphi \partial_\varphi  )
\ee
where $(\hat e_\eta  ,\hat e_\xi ,\hat e_\varphi )$ form an orthonormal
frame in $\RR^3$. Assuming now that the scalar field $\Phi$ only depends
on $\eta$, the energy functional simplifies to
\be \label{E-eta}
E[\Phi ] = 4\pi^2 \int_0^\infty d\eta \sinh \eta |\Phi_\eta |^3
\ee
where $\Phi_\eta \equiv \partial_\eta \Phi$. 

Remark: Due to the conformal base space symmetry of the Euler--Lagrange
equations, these equations are compatible with the ansatz $\Phi =\Phi (\eta
)$. It then follows from the principle of symmetric criticality that 
the resulting ODE
is identical to the Euler--Lagrange equation which is derived from
the reduced energy functional (\ref{E-eta}), see, e.g., \cite{Mak,mansut}.   

We prefer to introduce the new 
variable
\be
t = \sinh \eta
\ee
which leads to the energy functional
\be
E[\Phi ]= 4 \pi^2 \int_0^\infty dt t (1+t^2 ) |\Phi_t |^3.
\ee
We now skip the absolute value signs by assuming that $\Phi_t \ge 0
\, \forall \, t$, so that 
an infinitesimal variation of the scalar field leads to
\br 
\delta E &\equiv & 
E[ \Phi +\delta \Phi ] - E[\Phi ] \nonumber \\
&\simeq & 12\pi^2 \int_0^\infty
dt t (1+t^2) \Phi_t^2 \delta \Phi_t \nonumber \\ \label{bound1}
&=&  12 \pi^2 \left[ t(1+t^2) \Phi_t^2 \delta \Phi \right]_0^\infty -\\
&& 12 \pi^2 \int_0^\infty dt \delta \Phi \frac{d}{dt} [t(1+t^2)\Phi_t^2 ] 
\er
and, therefore, to the Euler--Lagrange equation
\be
\frac{d}{dt} [t(1+t^2)\Phi_t^2 ] =0
\ee
with the solution
\be \label{sol-toy}
\Phi_t = C[t(1+t^2)]^{-\frac{1}{2}}
\ee
where $C$ is an integration constant. For $C > 0$ it holds indeed that
$\Phi_t \ge 0$. The scalar field itself, 
\be \label{toy-sol2}
\Phi (t)= C\int_0^t \frac{dt'}{\sqrt{t' (1+t'^2)}} +c' , 
\ee
can be expressed in terms of elliptic functions, but 
we do not need the explicit expression here. The energy of this field
configuration is finite
\be
E = 4 \pi^2 C^3 \int_0^\infty \frac{dt}{\sqrt{t(1+t^2)}} = 
32 C^3 \pi^\frac{3}{2} \Gamma^2 (\frac{5}{4})
\ee
(where $\Gamma (\cdot )$ is the Gamma function) and can take any positive
value
due to the arbitrary integration constant $C$. Further, for this field
configuration the variation in the energy steming from the boundary
term (\ref{bound1}) is
\be
\delta E = 12 \pi^2 C^2 [  \delta \Phi (\infty ) - \delta \Phi (0)].
\ee
Specifically, for a variation proportional to the field,
$\delta \Phi =\epsilon \Phi$, this is nonzero,
\be
\delta E = 12 \pi^2 C^2 \epsilon [  \Phi (\infty ) -  \Phi (0)] =
96 C^2 \epsilon \pi^\frac{3}{2} \Gamma^2 (\frac{5}{4}).
\ee
In short, there exist finite energy solutions to the Euler--Lagrange equations
in spite of the scaling instability under $\Phi \to \lambda \Phi$, because
such a variation $\delta \Phi =\epsilon \Phi$ produces a nonzero contribution
to $\delta E$ from the boundary term.
It follows that the solution (\ref{sol-toy}), despite being of finite energy,
is {\em not} a critical point of
the energy functional. The instability is in agreement with more general
mathematical criteria for models without topological charges obtained by
Rubakov and others, see, e.g., page 58 of Ref. \cite{Mak}. 

\subsection{The AFZ model}
The degrees of freedom for both the AFZ and the Nicole model are 
given by a three-component unit vector field, that is, a map
\be
\vec n ({\bf r}) \, : \, \RR^3 \to S^2 \, ,\quad 
\vec n^2 =1 
\ee
where the tip of the unit vector field spans the unit two-sphere, or via
stereographic projection
\be
{\vec n} = {1\o {1+\mid u\mid^2}} \, ( u+\bar u , -i ( u-\bar u ) ,  
1-u\bar u ) \; ;
\qquad
u  = \frac{n_1 + i n_2}{1 + n_3}.
\label{stereo}
\ee
by a complex field
\be
u({\bf r}) \, : \, \RR^3 \to \CC_0 
\ee
where our conventions are such that the projection is from the south pole to 
the equatorial plane of the two-sphere. Further, $\CC_0$ is the one-point
compactified complex plane.

Topological solitons, i.e., static finite energy solutions with a non-zero 
integer value of the corresponding topological charge (the Hopf index)
have been found for both models, so let us briefly describe the
corresponding topological map (Hopf map). A Hopf map is a map
$\phi : S^3 \to S^2$ or, via the stereographic projection from $S^3$ to 
one-point compactified $\RR_0^3$, a map $\phi :\RR_0^3 \to S^2$, and is
characterized by the integer Hopf index (for details we refer to the Appendix 
A of Ref. \cite{ASGVW}). Therefore, the above fields $\vec n$ and $u$ can be
identified with Hopf maps provided that the following two conditions hold.
Firstly, the fields have to obey
\be
\lim_{|{\bf r}| \to \infty} \vec n =\vec n_0 = {\rm const}
\, ,\quad \lim_{|{\bf r}| \to \infty} u = u_0 = {\rm const}
\ee
so that they are, in fact, defined on one-point compactified 
Euclidean three-dimensional space $\RR_0^3$. Secondly, the fields have
to take values in the full target spaces $S^2$ and $\CC_0$, 
respectively, in such a way that the preimage of each point in target
space is a closed line in the base space $\RR_0^3$.

Both the AFZ and the Nicole model have a conformally invariant energy 
functional and conformally symmetric Euler--Lagrange equations, therefore
the Euler--Lagrange equations are, again, compatible with the ansatz
\be \label{tor-ans}
u= f(\eta ) e^{in\xi +im\varphi }
\ee
using toroidal coordinates. The limit $|{\bf r}| \to \infty$ corresponds
to the limits $\xi \to 0$ and $\eta \to 0$, therefore $f$ has to obey
$\lim_{\eta \to 0} f(\eta ) =0$ or  $\lim_{\eta \to 0} f(\eta ) =\infty $
in order that $u$ be defined on $\RR_0^3$. As $u$ and $1/u$ are related
by a symmetry transformation in both models, we may asssume for a true Hopf
map that $f(\eta =0) =0$ without loss of generality. For $u$ to take values
in the whole target space $\CC_0$ now requires that $f$  takes values on 
the whole positive real semiaxis $\RR_0^+$. Further, $f$ has to take the value
$\infty$ at $\eta =\infty$, because the pre-image of $\eta =\infty$ is a
circle (the unit circle in the $(x,y)$-plane), whereas the preimages of
finite $\eta =$ const. are surfaces (tori). So if we want the preimage of 
the point $u=\infty$ to be a circle, $u$ has to take the value $\infty$
at $\eta = \infty$.  In short, for a $u$ within the ansatz (\ref{tor-ans})
to be a genuine Hopf map, $f$ has to obey 
\be \label{hopf-bound}
f(\eta =0) =0 \, ,\quad f(\eta =\infty )=\infty \qquad \mbox{or}
\qquad f(t =0) =0 \, ,\quad f(t =\infty )=\infty
\ee
where $t=\sinh \eta$, as above. In Appendix A we derive the same result from
an analytic expression for the Hopf index by proving that integer valuedness
of the Hopf index precisely requires the boundary conditions 
(\ref{hopf-bound}).

The energy functional for the AFZ model is
\be
E_{\rm AFZ} [u]= \int d^3 {\bf r} \left( 
\frac{(\nabla u \cdot \nabla \bar u )^2 -
(\nabla u )^2 (\nabla \bar u )^2}{(1+u\bar u)^4} \right)^\frac{3}{4}
\ee
and, for the ansatz (\ref{tor-ans}) and using symmetric criticality,
\be
E_{\rm AFZ} [f]= 4\pi^2 \int_0^\infty d\eta \sinh \eta \left(
\frac{ 4f^2 f_\eta^2 (n^2 +
\frac{m^2}{\sinh^2 \eta })}{(1+f^2)^4}\right)^\frac{3}{4} .
\ee
For reasons of simplicity we now restrict to the case $m=n$ (the discussion 
for $n \not= m$ is completely analogous). Further we define $F\equiv f^2$
and find for the energy functional
\be
E_{\rm AFZ} [F]= 4\pi^2 m^\frac{3}{2} \int_0^\infty  
dt \frac{1+t^2}{\sqrt{t}} \left( \frac{F_t}{(1+F)^2}\right)^\frac{3}{2}
\ee
The solution of the Euler--Lagrange equation which obeys the boundary
condition $F(t=0)=0$, $F(t=\infty )=\infty$ required for a Hopf map is
$F=t^2$, see below. However, field configurations which deviate from these
boundary  conditions give perfectly valid, nonsingular 
energy densities with finite energy.
E.g., for $F=c_0 +t^2$, $c_0 \ge 0$, the energy is
\be
E_{\rm AFZ} (c_0)= 4\pi^2 m^\frac{3}{2}
\int_0^\infty dt t(1+t^2) (1+c_0 +t^2)^{-3}
=\pi^2 (2m)^\frac{3}{2} \frac{2+c_0}{(1+c_0)^2}.
\ee
Exactly the same result is obtained for $F=t^2/(1+c_0 t^2)$, which deviates 
from the boundary condition for a Hopf map at $t=\infty$.

Next, we want to find the Euler--Lagrange equation. We introduce the function
\be \label{G-F-rel}
G\equiv -\frac{1}{1+F} =-\frac{1}{1+f^2}
\ee
which leads to the energy functional
\be \label{func-G}
E_{\rm AFZ} [G] = 4\pi^2 m^\frac{3}{2} \int_0^\infty  
dt \frac{1+t^2}{\sqrt{t}} |G_t |^\frac{3}{2}
\ee
which is quite similar to the energy functional of the toy model in the
preceding section. As the energy functional is homogeneous in $G$, we again 
have that the target space scale transformation 
\be \label{G-scale}
G\to \lambda G
\ee 
is a
symmetry of the Euler--Lagrange equation which does not leave invariant
the energy. 

Remark: as long as we only consider the energy functional (\ref{func-G}) by
itself, $\lambda$ may take arbitrary values. However, the relation 
(\ref{G-F-rel}) requires, for positive semidefinite $F$, that 
$-G\in [0,1]$ and, therefore, restricts the possible values of
$\lambda $ to $0\le \lambda \le 1$.
 
For the variation of the energy functional
we easily find (assuming again $G_t \ge 0 \, \forall \, t$)
\br 
\delta E_{\rm AFZ} &\equiv & 
E_{\rm AFZ}[ G +\delta G ] - E_{\rm AFZ}[G ] \nonumber \\ \label{bound2}
&=&  6 \pi^2 m^\frac{3}{2} \left[ \frac{1+t^2}{\sqrt{t}} G_t^\frac{1}{2} 
\delta G \right]_0^\infty -\\
&& 6 \pi^2 m^\frac{3}{2} \int_0^\infty dt \delta G \frac{d}{dt} 
[\frac{1+t^2}{\sqrt{t}} G_t^\frac{1}{2} ] 
\er
and therefore the Euler--Lagrange equation
\be
\frac{d}{dt} 
[\frac{1+t^2}{\sqrt{t}} G_t^\frac{1}{2} ] =0
\ee
with the solution
\br
G_t &=& \frac{C}{2}\frac{t}{(1+t^2)^2} \\
G &=& -\frac{C}{1+t^2} -D
\er
where $C$ and $D$ are non-negative real integration constants. Further,
\be \label{F-sol}
F=\frac{1-C-D +t^2(1-D)}{C+D(1+t^2)}
\ee
is positive semidefinite by construction, therefore $C$ and $D$ are
restricted to
\be
C+D \le 1 \qquad \wedge \qquad D \le 1 .
\ee
For $C=1$, $D=0$ we find the solution $F=t^2$ which obeys the boundary 
conditions for a Hopf map. For other values of $C$ and $D$ we find 
solutions $F$ which do {\em not} obey these boundary conditions and,
therefore, do not give rise to Hopf maps. Nevertheless, they are
regular configurations which have finite energy. Specifically, for
$C$ strictly less than one the energy is lowered, 
\be
E_{\rm AFZ} 
= \frac{\sqrt{2}}{2} \pi^2 m^\frac{3}{2} C^\frac{3}{2} .
\ee 
Again, for variations proportional to the solution, $\delta G =\epsilon G$,
the boundary term (\ref{bound2}) gives a nonzero contribution to the
variation of the energy,
\be
\delta E_{\rm AFZ} = 3\sqrt{2} \pi^2 m^\frac{3}{2} \epsilon C^\frac{3}{2} .
\ee
Therefore, the solution to the Euler--Lagrange equation is again not a
critical point of the energy functional.

At this point we may ask to which symmetry transformation corresponds the 
target space scale transformation $G\to \lambda G$ in 
terms of the original fields $u$ and $\bar u$. This transformation is,
in fact, already well-known (the infinitesimal version has been 
found in \cite{BF}, and the finite version has been calculated in
\cite{ASG1}),
and it does not depend on the specific choice of
toroidal coordinates, but rather is a pure target space transformation.
It is based on the observation that the action density ${\cal L}_4$ is
just the square of the pullback (under the map $u$) of the area
twoform on the target space $S^2$,
\be \label{s2-ar}
\Omega =-i\frac{dud\bar u}{(1+u \bar u)^2} .
\ee
The transformation we search is the transformation $u\to v(u ,\bar u)$
such that the area twoform is mapped to
\be \label{BF-eq}
\frac{dud\bar u}{(1+\bar u u)^2} \ra \frac{dvd\bar v}{(1+\bar vv)^2}
=\Lambda^2 \frac{dud\bar u}{(1+\bar u u)^2}.
\ee
If we introduce the real coordinates on target space $u=F^{1/2}
e^{i\sigma}$ (angle and radius squared on the Euclidean plane;
here we do {\em not} assume a specific variable dependence of
the modulus squared $F$) and assume that
$v=(\tilde F)^{1/2}(F)e^{i\sigma}$ (i.e. $u$ and $v$ have the same argument,
and the transformation only affects the modulus) then we get the
equation
\be
\frac{\tilde F '(F)dFd\sigma}{(1+\tilde F)^2}=\Lambda^2 \frac{dFd\sigma}{
(1+F)^2}
\ee
or
\be
\frac{\tilde F'}{(1+\tilde F)^2} =\frac{\Lambda^2}{(1+F)^2}
\ee
with the solution
\be \label{int-const}
\frac{1}{1+\tilde F}=\frac{\Lambda^2}{1+F} +c
\ee
where $c$ is a constant of integration.
For $c=0$ we just get $\tilde G =\Lambda^2 G$, that is, the scale
transformation (\ref{G-scale}) with $\lambda =\Lambda^2$. 
 If, instead, 
we impose the boundary condition that $v=0$ for $u=0$ then 
$c=1-\Lambda^2$ and the solution
is
\be
\tilde F=\frac{\Lambda^2 F}{\Lambda^2 +(1+F)(1-\Lambda^2)}
\ee
or
\be \label{TS-transf}
v=\frac{\Lambda u}{[\Lambda^2 +(1+\bar uu)(1-\Lambda^2)]^{1/2}} .
\ee
Observe that $\Lambda$, again,  is restricted
to $\Lambda \le 1$ if $u$ is a genuine Hopf map which covers the whole
target space. Further, $v$ is no longer a true Hopf map for 
$\Lambda <1$.

Remark:  the transformation (\ref{TS-transf})  is a target space
symmetry transformation of the field equations, i.e., a target space
transformation which maps solutions $u$ into solutions $v$. It is, however,
not a symmetry of the action, but, instead,  
rescales the action 
density ${\cal L}_4$ by ${\cal L}_4 \to \Lambda^4 {\cal L}_4$. 
It is, therefore, not a Noether symmetry and does not define a Noether 
current and conserved charge. On the other hand, a combination of the
target space transformation (\ref{TS-transf}) and the base space
scale transformation (dilatation) $(t,{\bf r})\to (\mu t,\mu {\bf r})$,
$\mu =\Lambda^{-4}$, leaves the action invariant and, therefore, gives rise to
a Noether current and conserved charge, see \cite{BF} for details.

The transformation (\ref{TS-transf}) is a pure target space transformation
and, therefore, the rescaling of the action 
density ${\cal L}_4$ by ${\cal L}_4 \to \Lambda^4 {\cal L}_4$ is
independent
of the base space. As a consequence, stability problems which are
similar to the ones discussed here can be expected
for all lagrangian densities which are homogeneous in  ${\cal L}_4$,
for arbitrary base spaces. This issue will be discussed further in the 
last section.

\subsection{The Nicole model}

Here we just want to demonstrate briefly that, at least for field
configurations belonging to the ansatz (\ref{tor-ans}), soliton solutions 
of the Nicole model with
non-zero Hopf index are indeed separated by an infinite energy barrier from
the trivial sector, and that field configurations which do not obey the
boundary conditions necessary for an integer Hopf index have infinite
energy. 
The energy functional of the Nicole model is just a non-integer power
of the $CP^1$ model Lagrangian,
\be 
E_{\rm Ni}[u ,\bar u]= \int d^3 {\bf r} \left( \frac{\nabla u \cdot
\nabla \bar u}{(1+u\bar u)^2 }\right)^\frac{3}{2} .
\ee
For the ansatz (\ref{tor-ans}), and for $m=n$, which again we assume for
simplicity, this reduces to
\be
E_{\rm Ni}[f]=\int_0^\infty dt t(1+t^2) (1+f^2)^{-3} \left( f_t^2 +
m^2 \frac{f^2}{t^2} \right)^\frac{3}{2} \equiv \int_0^\infty dt
{\cal E}(t) .
\ee
Now let us assume that $f(0)=c_0>0$. 
For the density $ {\cal E}(t)$ this implies
\be
\lim _{t\to 0}{\cal E}(t) \simeq \lim _{t\to 0} t (1+c_0^2)^{-3}  
\left( f_t^2 + m^2 \frac{c_0^2}{t^2} \right)^\frac{3}{2}
 \ge t^{-2} \frac{m^3 c_0^3}{(1+c_0^2)^3}
\ee
which has a nonintegrable singularity at $t=0$. Equivalently, in the
limit $t\to \infty$ we have, assuming $f(\infty)=c_\infty$,
\be
\lim _{t\to \infty }{\cal E}(t) \simeq t^3 (1+c_\infty^2)^{-3} 
\left( f_t^2 + m^2 \frac{c_\infty^2}{t^2} \right)^\frac{3}{2}
\ge  \frac{m^3 c_0^3}{(1+c_0^2)^3} .
\ee
Therefore, the density is bounded from below by a nonzero constant, and its 
integral again will be infinite. 

It follows that for field configurations which obey boundary conditions such
that the corresponding Hopf index is non-integer, the energy in the Nicole
model automatically is infinite. In other words, it is a necessary
condition for a finite energy field configuration to have integer Hopf
index. The condition is not sufficient, however. For fields with the
leading behaviour $\lim _{t\to 0}{\cal E}(t) \simeq t^{\alpha_0}$, 
$\lim _{t\to \infty}{\cal E}(t) \simeq t^{\alpha_\infty}$, for instance,
it may be checked
easily that finite energy requires $\alpha_0 >1/3$ and $\alpha_\infty >1/3$,
which is more restrictive than just $f(0)=0$ and $f(\infty )=\infty $.
 
\section{Full energy functional and nonzero fluxes}
\subsection{Toy model}
The variation of the energy functional of the toy model for an arbitrary
scalar field is
\br
\delta E & \equiv & E[\Phi + \delta \Phi ] - E[\Phi ] \nonumber \\
&=& 3 \int d^3 {\bf r} \nabla \cdot \left( \delta \Phi 
|\nabla \Phi | \nabla \Phi \right) \\
&-&   3 \int d^3 {\bf r} \delta \Phi \nabla \cdot \left( 
|\nabla \Phi | \nabla \Phi \right)
\er
and the resulting Euler--Lagrange equation is just the conservation
equation
\be \label{cons-toy}
\nabla \cdot \left( 
|\nabla \Phi | \nabla \Phi \right) \equiv \nabla \cdot {\bf J} =0.
\ee
The nonzero contribution of the solution (\ref{toy-sol2}) 
to the boundary terms 
of the variation $\delta E$ of the reduced energy functional indicates
the presence of nonzero delta-function like source terms in the
conservation equation (\ref{cons-toy}). Further, these source terms
we expect to be concentrated at the loci of the boundaries of the
reduced system, that is, at $\eta =0$ (the $z$ axis) and at $\eta =\infty$
( the unit circle in the $(x,y)$-plane). This we want to investigate in the
sequel. For the solution $\Phi (\eta)$ of the last section with 
$\Phi_\eta =C \sinh^{-\frac{1}{2}} \eta$ the current ${\bf J}$ is
\be
{\bf J} = q^2 \Phi_\eta^2 \hat e_\eta =C^2 \frac{q^2}{\sinh \eta } 
\hat e_\eta .
\ee
We now want to calculate the flux of this current through a torus $\eta
=$ const. The surface element of a torus $\eta =$ const. is just
\be
d{\bf \Sigma}_{\rm T} = \frac{\sinh \eta}{q^2} \hat e_\eta d\xi d\varphi
\ee
therefore the flux simply is
\be  
\int d{\bf \Sigma}_{\rm T} \cdot {\bf J} =
\int_0^{2\pi} d\xi \int_0^{2\pi} d\varphi C^2 = 4\pi^2 C^2 .
\ee
So there is some nonzero total flux emerging from the $z$ axis and 
streaming towards the unit circle 
\be
{\cal C}=\{\vec x \in I\hspace{-0.12cm} R^3 :\;  z=0 \;
\wedge \; r^2 =1\}.
\ee
The total flux
escaping to infinity is zero, as may be checked by explicit calculation.  
However, this already follows from the fact that the total charges
distributed along the $z$ axis and along the unit circle ${\cal C}$ are
equal in magnitude and opposite in sign. The line charge density along the
circle ${\cal C}$ is constant, as is obvious for symmetry reasons. The
line charge density along the $z$ axis
may be calculated by calculating the flux through
a cylinder around the $z$ axis with infinitesimally small radius. For
infinitesimally small radius, the top and bottom of the cylinder do
not contribute. The surface element of the cylinder mantle in
cylinder coordinates $(\rho ,\varphi ,z)$, $\rho^2 =x^2 +y^2$, is
\be
d{\bf \Sigma}_{\rm M} = \rho \hat e_\rho d\varphi dz
\ee
Further, in the limit $\rho \to 0$, $\hat e_\eta \simeq \hat e_\rho $, 
and therefore
\br
\int d{\bf \Sigma}_{\rm M}\cdot {\bf J}
&=& \lim_{\rho \to 0} C^2 \int_{z_1}^{z_2} dz \int_0^{2\pi} d\varphi \rho
\frac{q^2}{\sinh \eta } \no \\ &=&
2\pi C^2 \int_{z_1}^{z_2} dz \frac{2}{1+z^2}
\er
where we used
\be
\rho^2 =\frac{\sinh^2 \eta}{q^2}
\ee
and 
\be
\sinh^2 \eta =\frac{4\rho^2}{4z^2 + (\rho^2 +z^2 -1)^2} \quad
\Rightarrow \quad \lim_{\rho \to 0} \frac{\sinh \eta}{\rho} = \frac{2}{1+z^2} .
\ee
Consequently, the solution $\Phi (\eta )$ solves in fact the inhomogeneous 
equation
\be \label{calQ}
\nabla \cdot {\bf J} =4\pi C^2  {\cal Q} \equiv 4\pi C^2
\left( \delta (x) \delta
(y) \frac{1}{1+z^2} - \delta (z) \delta (\rho^2 -1) \right) ,
\ee
and the total charge of the density distribution ${\cal Q}$ is zero,
\be \label{zero-ch}
\int d^3{\rm r} 4\pi C^2  {\cal Q} =\int d^3{\rm r}\nabla \cdot {\bf J}=0 , 
\ee
as may be checked easily.  

Remark: 
The observation  
that the toy model is equivalent to the purely electrostatic sector of a 
nonlinear theory of electrodynamics may add some additional interest to it.
Indeed, take the Lagrangian density
\be
{\cal L} = -\frac{1}{4} F_{\mu \nu}F^{\mu \nu}
\left( \frac{ F_{\alpha\beta}F^{\alpha\beta}}{ 2\Omega^4} \right)^\frac{1}{2}
\ee
where $\Omega$ is a constant with the dimension of mass, and
$F_{\mu\nu}=\partial_\mu A_\nu - \partial_\nu A_\mu$, $A_\mu =(V, \vec A)$. 
This leads to the field equation (Gauss law)
\be
\partial_\nu \left[ F^{\mu \nu}
\left( \frac{ F_{\alpha\beta}F^{\alpha\beta}}{ 2\Omega^4} \right)^\frac{1}{2}
\right] =0
\ee
or, in the electrostatic case ($\vec A=0$, $V=V({\bf r})$, $\vec E=-\nabla
V$), to 
\be \label{cons-nled}
\nabla \cdot \left( 
|\nabla V | \nabla V \right)  =0
\ee
which is identical to Eq. (\ref{cons-toy}) when the potential
$V$ is identified with the scalar field $\Phi$ of the toy model. Therefore, we
have the electrostatic solutions
\be
V_t =\frac{C}{\sqrt{t(1+t^2)}}
\ee
and
\be
\vec E = -qV_\eta \hat e_\eta \equiv -q\sqrt{1+t^2} V_t \hat e_\eta
=-q\frac{C}{\sqrt{t}} \hat e_\eta .
\ee
The electric induction in this model
\be
\vec D=\frac{3q^2}{2\Omega^2}|\vec E| \vec E
\ee
is proportional to the current ${\bf J}$ of the toy model and, therefore,
obeys the inhomogeneous divergence equation
\be
\nabla \cdot \vec D = -\frac{3C^2}{2\Omega^2}{\cal Q}
\ee
where the singular line
charge density ${\cal Q}$ is defined in Eq. (\ref{calQ}).

\subsection{AFZ model}
The variation of the energy functional of the AFZ model is
\br
\delta E_{\rm AFZ} & \equiv & E[u, \bar u + \delta \bar u ] - 
E[u,\bar u ] \nonumber \\
&=&  \frac{3}{2} \int d^3 {\bf r} \nabla \cdot \left( \delta \bar u
(1+u\bar u )^{-2} {\bf L} \right) \\
&-&  \frac{3}{2}  \int d^3 {\bf r} \delta \bar u (1+u\bar u )^{-2} 
\nabla \cdot {\bf L}
\er
where 
\br
{\bf L} &\equiv & (1+u\bar u)^{-1} H^{-\frac{1}{4} } {\bf K} \label{L-cur} \\
H &\equiv & (\nabla u \cdot \nabla \bar u )^2 - (\nabla u)^2 
(\nabla \bar u )^2 \\
{\bf K} &\equiv &  (\nabla u \cdot \nabla \bar u ) \nabla u -
(\nabla u)^2 \nabla \bar u .
\er
The resulting Euler--Lagrange equation is the conservation
equation
\be \label{cons-afz}
\nabla \cdot {\bf L} =0.
\ee
We want to investigate whether, again, there are delta-function like
source terms present at the r.h.s. of this conservation equation.
However, for the AFZ model things are complicated by the fact that,
due to the infinitely many symmetries of the model, there exist in fact
infinitely many currents such that their conservation is equivalent to the
Euler--Lagrange equations. Indeed, with the help of the identity
\be
\nabla u \cdot  {\bf K} \equiv 0
\ee
it easily follows that together with the conservation equation
(\ref{cons-afz}) we have the infinitely many conservation equations
\be \label{cons-afz-z}
\nabla \cdot {\bf L}^\zeta \equiv \nabla \cdot \zeta {\bf L}  =0
\ee
where $\zeta =\zeta (u)$ is an arbitrary function of $u$ only. The currents
$  {\bf L}^\zeta $ are, in general, complex, but it is easy to construct
an equivalent set of real conserved currents,
\be
{\bf L}^{\cal G} \equiv -i \left( {\cal G}_u {\bf L} - {\cal G}_{\bar u} 
\bar {\bf L} \right)
\ee 
where ${\cal G}_u \equiv \partial_u {\cal G}$, etc., and ${\cal G}
={\cal G}(u,\bar u)$ is an arbitrary {\em real} function of $u$ and $\bar u$.
The real currents ${\bf L}^{\cal G}$ are just the Noether currents of the
area-preserving target space diffeomorphisms, which are wellknown
symmetries of the AFZ model \cite{BF,FeRa}.

For the ansatz (\ref{tor-ans}) for $u$ in toroidal coordinates (and $m=n$), 
the current (\ref{L-cur}) is
\be
{\bf L} = 2\frac{q^2}{1+f^2} e^{im(\xi + \varphi )} \left[ \left(
\frac{m}{\tanh \eta} f \right)^\frac{3}{2} f_\eta^\frac{1}{2}
\hat e_\eta + i   \left( \frac{m}{\tanh \eta} f \right)^\frac{1}{2} 
f_\eta^\frac{3}{2} \left( \hat e_\xi + \frac{\hat e_\varphi
}{\sinh \eta } \right) \right] .
\ee
This current will give a zero total flux through a torus $\eta =$ const. 
because of the angular factor $\exp [im(\xi + \varphi )]$. By inspection,
it is obvious that a nonzero flux will exist for the current ${\bf L}^\zeta $
with $\zeta = u^{-1}$, because then the angular factor is absent.
Multiplying by the surface element of a torus $\eta =$ const. we find
\br
{\bf L}^{\frac{1}{u}} \cdot d{\bf \Sigma}_{\rm T} &=& 2m^\frac{3}{2}
\frac{\cosh^\frac{3}{2}}{\sinh^\frac{1}{2}\eta } 
\left( \frac{ff_\eta}{(1+f^2)^2}
\right)^\frac{1}{2} \nonumber \\
&=&  \sqrt{2} m^\frac{3}{2}
\frac{1+t^2}{\sqrt{t}} \left( \frac{F_t}{(1+F)^2}
\right)^\frac{1}{2} \nonumber \\
&=&  C^\frac{1}{2} m^\frac{3}{2}
\er
where we used the solution
\be \label{sol2-G_t}
\frac{F_t}{(1+F)^2} \equiv G_t = \frac{C}{2}\frac{t}{(1+t^2)^2} \,\, ,
\ee
and therefore the flux $ 4\pi^2 m^\frac{3}{2}\sqrt{C}$ through an arbitrary
torus of $\eta =$ const.

Remark: the current with nonzero flux may be equally found among the real
currents ${\bf L}^{\cal G}$ for the choice
\be
{\cal G} \equiv {\cal G}^{\rm f} = \frac{i}{2} (\ln u - \ln \bar u ).
\ee
The current ${\bf L}^{{\cal G}^{\rm f}}$ 
is identical to the current ${\bf J}$ of the
toy model and, therefore, obeys the inhomogeneous divergence equation
\be \label{div-inh-L}
\nabla \cdot {\bf L}^{{\cal G}^{\rm f}} = 4\pi C^\frac{1}{2}  m^\frac{3}{2}
\left( \delta (x) \delta
(y) \frac{1}{1+z^2} - \delta (z) \delta (\rho^2 -1) \right) .
\ee

\section{Discussion}
The main result of this paper is the observation that in the AFZ model
for any solution $u$ there exists a one-parameter family of solutions
$v(u,\Lambda )$ which connects the given solution $u$ at $\Lambda =1$ to
the vacuum solution $v(u,0)=0$, see Eq. (\ref{TS-transf}). 
Therefore, sectors of field configurations
with different topological
(Hopf) index are not separated by an infinite energy barrier, and there exist
well-defined (i.e. single-valued, nonsingular) field configurations with 
finite energy in the model which do not correspond to integer Hopf index.
For static solutions (i.e., solitons), this observation immediately implies 
some problems. The first problem is that the mere existence of soliton 
solutions under these conditions poses an apparent paradox. Indeed, the
static field equations which the solitons obey are just the Euler-Lagrange
equations for the variational problem of the energy functional
$E_{\rm AFZ}[u,\bar u]$ and, therefore, if $u_0$ is a soliton solution, it
should hold that 
\be
\left[ \frac{d}{d\epsilon} E_{\rm AFZ}[u_0,\bar u_0 +\epsilon \delta \bar u ] 
\right]_{\epsilon =0} =
\left[ \frac{d}{d\epsilon} E_{\rm AFZ}[u_0 + \epsilon \delta u,\bar u_0 ] 
\right]_{\epsilon =0} =0
\ee 
for arbitrary $\delta u$. But in our case this is not true. We have that
$ E_{\rm AFZ}[v,\bar v] = \Lambda^3  E_{\rm AFZ}[u,\bar u]$ and therefore
\be
\left[ \frac{d}{d\epsilon} E_{\rm AFZ}[u_0,\bar u_0 +\epsilon \delta \bar u ] 
\right]_{\epsilon =0} +
\left[ \frac{d}{d\epsilon} E_{\rm AFZ}[u_0 + \epsilon \delta u,\bar u_0 ] 
\right]_{\epsilon =0} = 3E_{\rm AFZ}[u_0,\bar u_0]
\ee 
for $\Lambda =1-\epsilon$ and $\delta u = u_0(1+u_0\bar u_0 )$. The
resolution of this apparent paradox can be understood in different ways.
For the reduced energy functional within the ansatz in toroidal coordinates
(see Section 2),  it turned out that in addition to the
bulk contribution to the variation of the energy functional (which is 
zero for a solution) there are boundary contributions, which are nonzero
for certain variations, even when the variations are performed about a
soliton solution.

For the full energy functionals the nonzero contributions to the variation 
about a solution cannot be attributed to boundary terms, because there
are no boundaries in the base space $\RR^3$, and the soliton solutions are
behaving well in the limit ${\bf r}\to \infty$. 
For the toy model, the resolution
of the paradox is quite simple. The static field equation corresponds to
a conservation equation $\nabla \cdot {\bf J}=0$ for some current $ {\bf J}$,
whereas the ``solution'' obeys, in fact, an inhomogeneous conservation
equation instead, with some delta-function like sources and sinks
on the r.h.s.
For the AFZ model, the resolution is slightly more subtle. The static
field equation is again equivalent to a conservation equation
 $\nabla \cdot {\bf L}=0$, 
but now the well-known soliton solutions do not induce
an inhomogeneous source term in this conservation equation. However, due to
the infinitely many symmetries of the AFZ model, there formally exist 
infinitely many more conservation equations 
$\nabla \cdot {\bf L}^{\zeta}=0$ for 
soliton solutions. It turns out that, for nontrivial soliton solutions,
one of these infinitely many conservation equations is, in fact, 
inhomogeneous and contains, again, a singular flux term on the r.h.s., 
see Eq. (\ref{div-inh-L}).
Further, the strength of the delta-function like inhomogeneous term is 
not varied
in the variational derivation of the field equations. Therefore, the value
of $\Lambda$, which plays the role of an integration constant for
solutions within the ansatz in toroidal coordinates ($\Lambda =\sqrt{C}$,
see Eqs. (\ref{sol2-G_t}) and (\ref{div-inh-L})), remains fixed, which
explains the existence of solutions to the variational equations. 

The next issue to be discussed is, of course, the stability of
solutions (that is, the stability under small perturbations). 
Naivly, one might assume that the presence of the inhomogeneous
term in one of the conservation equations stabilizes the solutions by
fixing the value of $\Lambda$. However, this might not be correct for the 
following reasons. Firstly, the inhomogeneous term does not correspond to a
nonzero charge, because the spatial integral of the inhomogeneity is zero,
see Eq. (\ref{zero-ch}). Secondly, the fact that the inhomogeneous term
is not varied by the variation of the energy functional does not imply that
it is conserved under dynamical evolution of the system. The conservation
under dynamical evolution would require that $\Lambda$ (or a related
quantity, like, e.g., the Hopf charge, which varies under the transformation
(\ref{TS-transf})) may be expressed by one or more of the conserved
charges of the theory. For static solutions this most likely does not happen,
however, because most of the conserved charges are trivially zero for
static solutions (e.g., all the 
infinitely many conserved charges related to the
infinitely many area-preserving target space diffeomorphisms contain time
derivatives and are, therefore, zero for static solutions).
We conclude that the stability of the soliton solutions in the
AFZ model is problematic.

Although the explicit calculations have been done for the static AFZ model
with Lagrangian $({\cal L}_4)^\frac{3}{4}$ and base space $\RR^3$, some
conclusions may be drawn for more general models based on 
the Lagrangian ${\cal L}_4$ and with different base spaces. 
For instance, static soliton solutions have been found in \cite{Fer1}
for the Lagrangian ${\cal L}_4$ and the three-sphere $S^3$ as base space
(that is, space-time $S^3 \times \RR$). These static solutions solve
the Euler--Lagrange equations which follow from varying the static
energy functional 
\be
E = \int_{S^3} dV_{S^3} 
\frac{(\nabla u \cdot \nabla \bar u )^2 -
(\nabla u )^2 (\nabla \bar u )^2}{(1+u\bar u)^4} 
\ee
(where $dV_{S^3} $ is the volume element on $S^3$) and have finite energy, 
therefore the same arguments as above apply. Due to the symmetry 
(\ref{TS-transf}) these solutions are connected to the vacuum and,
in addition, are not critical points of the energy functional. Therefore,
the mere existence of these solutions requires the presence of
inhomogeneous singular flux terms in some (at least one)
of the infinitely many conservation equations. This line of arguments remains
valid for static finite energy solutions for Lagrangians which are 
arbitrary powers of the Lagrangian  ${\cal L}_4$, and for arbitrary base
spaces.

The situation is slightly different for time-dependent solutions. Indeed,
time dependent solutions for the Lagrangian  ${\cal L}_4$ have been found
both for base space-time $\RR^4$ and $S^3 \times \RR$ in \cite{Fer2} and
\cite{Fer3}, respectively. These solutions have integer Hopf charge and are
particle-like, that is, they have finite energy and infinite action.
These solutions are again connected to the vacuum by the symmetry 
transformation (\ref{TS-transf}), but the apparent paradox discussed above
is absent for these time dependent solutions. The energy transforms
like $E \to \Lambda^4 E$ under the transformation  (\ref{TS-transf}),
but the time dependent field equations are not derived by varying the energy 
functional, and time dependent solutions are not required to be critical
points of the energy functional. The action $I$ transforms like the energy
under the symmetry transformation  (\ref{TS-transf}), $I\to \Lambda^4 I$,
so the same apparent paradox would exist for time dependent finite action
solutions (``instantons''). But it does not exist for time dependent
solutions with infinite action (particle-like solutions). 
Therefore, the existence of inhomogeneous singular flux terms in 
some conservation equations is not required by the existence of these time
dependent solutions (although we do not know whether such inhomogeneous
singular flux terms do or do not show up for the time dependent solutions). 
Even the issue of stability may be less problematic for these time dependent 
solutions. The point is that the infinitely many conserved Noether
charges, which exist in the model as a consequence of its infinitely many
symmetries, are nontrivial (i.e., nonzero) for time dependent solutions.
It is quite possible that these infinitely many conservation laws
stabilize the time dependent solutions in spite of the fact that those
solutions 
are connected to the vacuum by a symmetry transformation. 

Another interesting feature of models based on the Lagrangian density
${\cal L}_4$, which is related to the existence of finite energy
configurations with non-integer Hopf index, is the rich vacuum structure
of such models. Indeed, a complex field $u(h)$ which only depends on one
real function  $h(t,{\bf r})$ gives zero when inserted into the
Lagrangian density ${\cal L}_4$, implying that the vacuum manifold is 
infinite dimensional. This may be understood in geometric terms by
the observation that the quartic Lagrangian ${\cal L}_4$ (and the related
energy density) is just the square of the pullback (under the map $u$)
of the area two-form on target space $S^2$. A field 
$u(h)$ of the above type is a map $\RR^3 \to {\cal M}^1 \to S^2$, where
${\cal M}^1$ is a one-dimensional manifold, and
the pullback of a two-form onto a one-dimensional manifold is zero.
As a consequence, static fields $u$ which have
a functional dependence $u(h)$ in the limit ${\bf r}\to \infty$ will 
have finite energy. In the AFZ model and within the ansatz of toroidal
coordinates, for instance, it happens that $\lim_{{\bf r}\to \infty}
u =c_0 \exp (im\varphi )$ whenever $f(\eta)$ obeys the boundary condition
$f(0)=c_0$.

The rigorous results, as well as the detailed arguments summarized in 
this section for
some nonlinear models in 4 dimensions, are the main contribution of our
work. These models are special, but they are related to and share many
properties of some relevant fundamental and effective field theories. 
The Lagrangian ${\cal L}_4$, for instance, is just the quartic part of
the Faddeev--Niemi model.
The stability of this term has been investigated recently in 
\cite{Speight}, together with some generalizations (Lagrangians based
on the Kaehler two-form on various target spaces). 
However, the stability analysis
of the quartic Lagrangian ${\cal L}_4$
in  \cite{Speight} focused on the quadratic
variation of the energy functional, whereas we found 
that already the linear variation is problematic. 
Our general conclusion, therefore, 
is that  the stability of the extended solutions, which is
often taken more or less for granted, is an extremely subtle and difficult
issue which, as shown here, can be 
understood in detail by a careful analysis in some cases.
Our methods can certainly be useful for that purpose. Note  also that in
numerical analysis the restriction to a fixed topological sector is
often imposed from the outset by assuming the 
corresponding boundary conditions.
Under these conditions, it would not be possible 
via numerical analysis to detect stability
problems of the type discussed here.

\section*{Appendix A}
Here we want to demonstrate briefly from the analytic expression for
the Hopf index  
that for fields $u$ within the toroidal
ansatz (\ref{tor-ans}) the condition that $u$ has integer Hopf index precisely
requires that $f$ obeys the boundary condition (\ref{hopf-bound}). 
The analytic expression for the Hopf index is
\be
Q=\frac{1}{16 \pi^2}\int d^3 r \vec {\cal A} \cdot \vec {\cal B}
\ee
where $\vec {\cal B}$ is the Hopf curvature
\be \label{ho-curv}
\vec {\cal B} =\frac{2}{i} \frac{\nabla u \times \nabla 
\bar u}{(1 + u\bar u)^2} 
\ee
and $\vec {\cal A}$ is the gauge potential for the ``magnetic field''
$\vec{\cal B}$, $\vec {\cal B} = \nabla \times \vec {\cal A}$. 
There is no local expression for $\vec {\cal A}$ in terms of the Hopf map
$u$ alone, but when $u$ is expressed with the help of a four component
unit vector field $e_\alpha$, $\alpha = 1, \ldots ,4$ like
\be
u=\frac{e_1 +ie_2}{e_3 +ie_4} ,
\ee
then an explicit, local expression for the gauge
potential $\vec {\cal A}$ in terms of the $e_\alpha$ exists and is given by
\be
\vec {\cal A} = \frac{2}{i} [(e_1 -ie_2 ) \nabla (e_1 +ie_2) +
(e_3 -ie_4 )\nabla (e_3 +ie_4 )] .
\ee
In our case we may choose
\be 
e_1 +ie_2 = \frac{f}{\sqrt{1+f^2}}e^{im\varphi} \, , \quad
e_3 +ie_4 = \frac{1}{\sqrt{1+f^2}}e^{-in\xi}
\ee
and, therefore,
\be
\vec {\cal A} = \frac{2}{i}q \frac{1}{1+f^2} \left( \frac{2ff_\eta}{1+f^2}
\hat e_\eta +\frac{imf^2}{\sinh \eta } \hat e_\varphi -in\hat e_\xi
\right) 
\ee
\be
\vec {\cal B} = 4q^2 \frac{ff_\eta}{(1+f^2)^2} \left( n\hat e_\varphi
- \frac{m}{\sinh \eta } \hat e_\xi \right)
\ee
and
\be
Q=\frac{1}{16 \pi^2}\int d\eta d\xi d\varphi 8mn \frac{ff_\eta}{(1+f^2)^2}
= mn\left[ -\frac{1}{1+F(\eta)} \right]_0^\infty
\ee
(remember $F\equiv f^2$). It follows that the conditions for integer, nonzero
Hopf index are precisely $f(0)=0$ and $f(\infty )=\infty$ (or the inverse
conditions $f(0)=\infty $ and $f(\infty )=0$; however, as $u\to (1/u)$ is
a symmetry transformation for both models, 
these cases are equivalent). \\ \\ \\
{\large\bf Acknowledgement:} \\
This research was partly supported by MCyT(Spain) and FEDER
(FPA2005-01963), Incentivos from Xunta de Galicia and the EC
network "EUCLID". CA acknowledges support from the
Austrian START award project FWF-Y-137-TEC and from the  FWF
project P161 05 NO 5 of N.J. Mauser. AW thanks the Physics Department of
the University of Santiago de Compostela for hospitality.
Further, the authors thank L. Ferreira for helpful discussions.


\begin{thebibliography}{99}
\bibitem{AFSG}
O. Alvarez, L.A. Ferreira, and
J. S\'{a}nchez-Guill\'{e}n, Nucl. Phys. B {\bf 529}, 689 (1998).
\bibitem{Mak}
V.G. Makhankov, Y.P. Rubakov, and V.J. Sanyuk, ``The Skyrme Model'',
Springer Verlag, Berlin 1993.
\bibitem{mansut}
N. Manton, P. Sutcliffe, ``Topological Solitons'', Cambridge University
Press, Cambridge, 2004.
\bibitem{Skyrme}
T.H.R. Skyrme, Proc. R. Soc. Lond. A260 (1961) 127; Nucl. Phys. 31 (1962) 556.
\bibitem{Witt}
E. Witten, Nucl. Phys. B223 (1983) 433.
\bibitem{HMS}
C.J. Houghton, N.S. Manton, P.M. Sutcliffe,
Nucl. Phys. B510 (1998) 507, hep-th/9705151.
\bibitem{BS1}
R. Battye, P. Sutcliffe,
Phys. Rev. C73 (2006) 055205,
hep-th/0602220
\bibitem{HoMa}
C. Houghton, S. Magee, 
hep-th/0602227.
\bibitem{Fad}
L.D. Faddeev, in ``40 Years in Mathematical Physics'', World Scientific, 
Singapore 1995.
\bibitem{FN1}
L.D. Faddeev and A.J. Niemi,
Nature 387 (1997) 58; hep-th/9610193.
\bibitem{LiYa}
F. Lin, Y. Yang, Comm. Math. Phys. {\rm 249}, 273 (2004).
\bibitem{GH}
J. Gladikowski and M. Hellmund,
Phys. Rev. D56 (1997) 5194; hep-th/9609035.
\bibitem{BS2}
R.A. Battye and P. Sutcliffe,
Proc. Roy. Soc. Lond. A455 (1999) 4305, hep-th/9811077.
\bibitem{BS3}
R.A. Battye and P. Sutcliffe, Phys. Rev. Lett. 81 (1998) 4798.
\bibitem{HiSa}
J. Hietarinta and P. Salo, Phys. Rev. D62 (2000) 081701.
\bibitem{AFZ1} 
H. Aratyn, L.A. Ferreira, and A. Zimerman, 
Phys. Lett. B456 (1999) 162.
\bibitem{AFZ2} 
H. Aratyn, L.A. Ferreira, and A. Zimerman, 
Phys. Rev. Lett. 83 (1999) 1723.
\bibitem{BF} 
O. Babelon and L.A. Ferreira, 
JHEP 0211 (2002) 020.
\bibitem{Ni} D.A. Nicole, 
J. Phys. G4 (1978) 1363.
\bibitem{Deser} 
S. Deser, M.J. Duff, C.J. Isham,
Nucl. Phys. B114 (1976) 29.
\bibitem{ASG2}
C. Adam and J. Sanchez-Guillen, JHEP01 (2005) 004. 
\bibitem{nicole_modif} 
A. Wereszczy\'{n}ski, Eur. Phys. J. C 41 (2005) 265.
\bibitem{ASGVW}
C. Adam,   J. S\'anchez-Guill\'en,  R.A. V\'azquez, and A. Wereszczy\'nski,
J. Math. Phys.  47, 052302 (2006),
hep-th/0602152.
\bibitem{fake}
M. Novello, E. Huguet, J. Queva, astro-ph/0604475 
\bibitem{MOND}
J.D. Bekenstein, PoS JHW2004 (2005) 012;
astro-ph/0412652
\bibitem{ASG1}
C. Adam and   J. S\'anchez-Guill\'en,
J. Math. Phys. 44, 5243 (2003),
hep-th/0302189.
\bibitem{FeRa}
L.A. Ferreira, A.V. Razumov,
Lett. Math. Phys. 55 (2001) 143;
hep-th/0012176
\bibitem{Fer1}
E. De Carli, L.A. Ferreira,
J.Math.Phys.46:012703,2005
hep-th/0406244
\bibitem{Fer2}
L.A. Ferreira,
JHEP 0603:075,2006
hep-th/0601235
\bibitem{Fer3}
A.C. Riserio do Bonfim, L.A. Ferreira,
JHEP 0603:097,2006
hep-th/0602234
\bibitem{Speight}
J.M. Speight, M. Svensson, math.DG/0605516.
\end{thebibliography}
\end{document}